\documentclass[pdftex,superscriptaddress,floats,showpacs]{revtex4}
\usepackage{latexsym}
\usepackage{amsfonts}
\usepackage{amssymb}
\usepackage{graphicx}
\begin{document}
\title{Laser-driven high-power X- and gamma-ray ultra-short pulse source}
\author{Timur Zh. Esirkepov}
\author{Sergei V. Bulanov}
\affiliation{Advanced Photon Research Center, Japan Atomic Energy Agency, 8-1 Umemidai, Kizugawa, Kyoto, 619-0215 Japan}
\author{Alexei G. Zhidkov}
\affiliation{Central Research Institute of Electric Power Industry, 2-6-1 Nagasaka, Yokosuka-shi, Kanagawa-Ken, 240-0196, Japan}
\author{Alexander S. Pirozhkov}
\author{Masaki Kando}
\begin{abstract}
A novel ultra-bright high-intensity source of X-ray and gamma radiation is suggested.
It is based on the double Doppler effect, where
a relativistic flying mirror reflects a counter-propagating
electromagnetic radiation causing its frequency multiplication and intensification,
and on the inverse double Doppler effect, where the mirror
acquires energy from an ultra-intense co-propagating electromagnetic wave.
The role of the flying mirror is played by a high-density thin plasma slab
accelerating in the radiation pressure dominant regime.
Frequencies of high harmonics generated at the flying mirror by
a relativistically strong counter-propagating radiation
undergo multiplication with the same factor as
the fundamental frequency of the reflected radiation,
approximately equal to the quadruple of the square of the mirror Lorentz factor.
\end{abstract}
\maketitle
\section{Introduction}
\label{sec:intro}

An electromagnetic wave reflected off a moving mirror
undergoes frequency multiplication and
corresponding increase in the electric field magnitude.
This phenomenon sometimes called the double Doppler effect
was discussed by A. Einstein in his seminal paper \cite{bib:Einstein}
where the frequency multiplication factor
was calculated as an example of the use of Lorentz transformations.
The multiplication factor is approximately
proportional to the square of the Lorentz factor of the mirror,
making this effect an attractive basis for a source
of powerful high-frequency radiation.
In relativistic plasma,
the double Doppler effect manifests itself
when a fast change of the electric current density
leads to the conversion of an incident light
into strongly compressed pulses
of high-frequency electromagnetic radiation.

Pulse compression and frequency upshifting
can be seen in a broad variety of configurations, \cite{bib:Rev}.
A specular reflection can be afforded by a sufficiently dense
relativistic electron cloud as suggested in Refs. \cite{bib:Landecker-52,bib:Ostr};
a less dense bunch of relativistic electrons
causes the backward Thomson scattering as discussed in Refs. \cite{bib:Arutyunian-63}.
The reflection at the moving ionization fronts was studied in Refs. \cite{bib:Semenova-67}.
Further examples of the manifestation of the double Doppler effect
in plasma whose dynamics is governed by the strong collective fields
are seen in the concepts of the sliding mirror \cite{bib:SM},
oscillating mirror \cite{bib:OscM,bib:FOIL} and flying mirror \cite{bib:FM}
which can produce ultra-short pulses of XUV radiation and X-ray.

The sliding mirror is formed by a thin foil whose density is so high that
the electrons are confined whithin the boundaries of the ion layer.
Irradiated by a relativistically strong laser pulse,
which is not capable to quickly break the confinement condition,
these electrons perform nonlinear motion along the foil,
enriching the (partially) reflected radiaiton (as well as transmitted radiation)
with high harmonics, \cite{bib:SM}.
In a less dense foil the electrons can perform collective motion in the direction
perpendicular to the foil, thus forming a mirror oscillating with relativistic
velocity.
A portion of an incident relativistically strong electromagnetic wave,
driving the oscillating mirror,
is reflected in the form of strongly distorted wave
carrying high harmonics, \cite{bib:OscM,bib:FOIL,bib:OscM2}.

In the flying mirror concept \cite{bib:FM},
the role of the mirror is played by the electron density
modulations in a strongly nonlinear Langmuir wave
excited by an intense laser pulse (driver) in its wake in underdense plasma.
A relatively weak counter-propagating electromagnetic wave (source)
is (partially) reflected at these modulations moving with the velocity
equal to the group velocity of the driving laser pulse.
In addition, due to a finite waist of the driver pulse,
the electron density modulations take a paraboloidal shape \cite{bib:parab},
and hence focus the reflected radiation (signal).
The most efficient reflection is afforded
by a breaking wake wave, where
the caustics of the plasma flow are formed and correspondingly
the electron density becomes formally singular.
The reflection coefficient is calculated in \cite{bib:Panchenko}
for a broad class of caustics.
For the case of the breaking Langmuir wave, the reflection efficiency
is high enough to acess, with present-day technology,
the quantum electrodynamics (QED) critical field
in the focus of the reflected signal \cite{bib:FM}.

Here we discuss a novel scheme of the flying mirror,
the accelerating double-sided mirror,
which can efficiently reflect
the counter-propagating relativistically strong electromagnetic radiation.
The role of the mirror is played
by a high-density plasma slab which is accelerated as a whole
by an ultra-intense laser pulse (the driver)
in the Radiation Pressure Dominant (RPD) regime
(synonymous to the Laser Piston regime), \cite{bib:RPD}.
Such an acceleration can be described as the mirroring effect:
it is the reflection that allows the energy transfer from
the driver radiation to the co-propagating plasma slab.
This effect is inverse with respect to the double Doppler effect.
This plasma slab also acts as a mirror
for a counter-propagating relativistically strong electromagnetic radiation (the source).
As such it exhibits the properties of the sliding and oscillating mirrors, producing high harmonics.
As a result, in the spectrum of the reflected radiation
both the fundamental frequency of the incident radiation and all the high harmonics
are multiplied by the same factor, approximately
proportional to the square of the Lorentz factor of the mirror.
This concept opens the way towards extremely bright
sources of ultrashort energetic bursts of X-ray and gamma-ray,
which become realizable with present-day technology.

In the next sections we recall some aspects of the flying mirror concept,
then we present the scheme of the accelerating mirror and corresponding computer simulations.
Finally, we discuss the prospects of the proposed concept.

%%%%%%%%%%%%%%%%%%%%%%%%%%%%%%%%%%%%%%%%%%%%%%%%%%%%%%%%%%
\section{The frequency multiplication factor}
\label{sec:freq}

\begin{figure}[th]
\centerline{\includegraphics[scale=0.4]{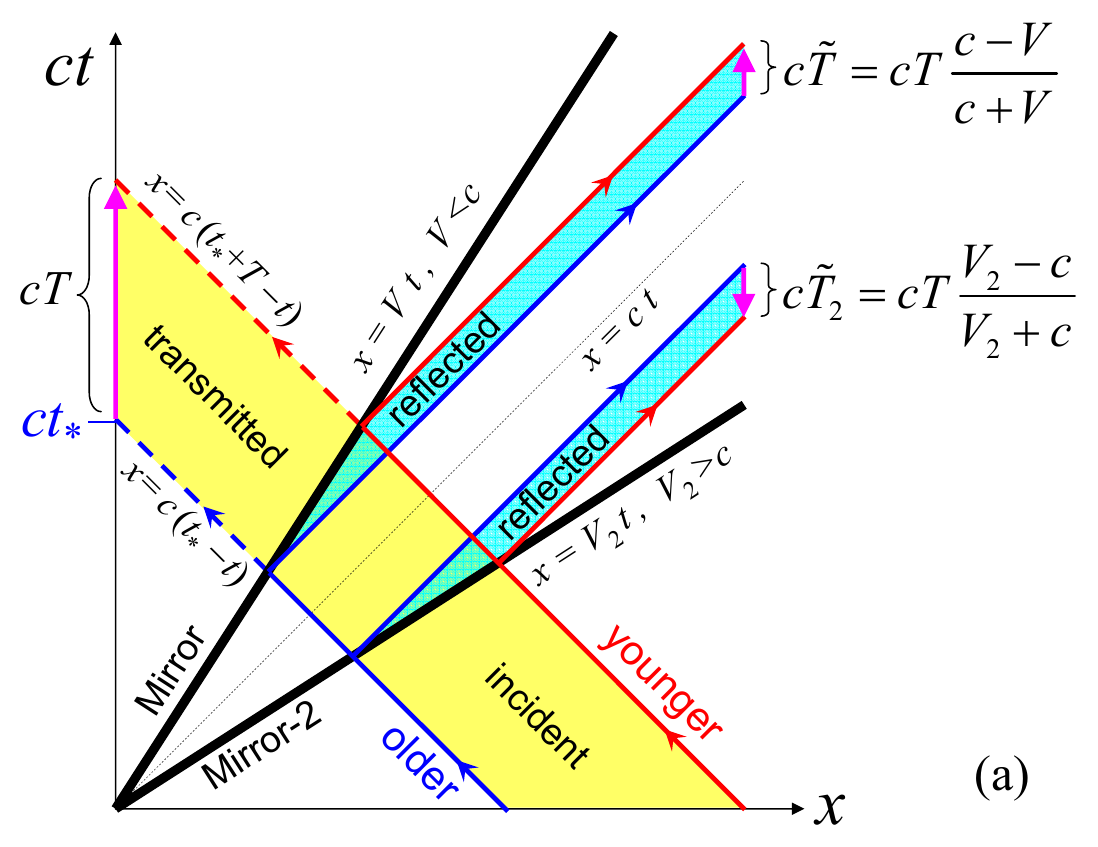}
\hspace*{8ex}
%\centerline{
\includegraphics[scale=0.4]{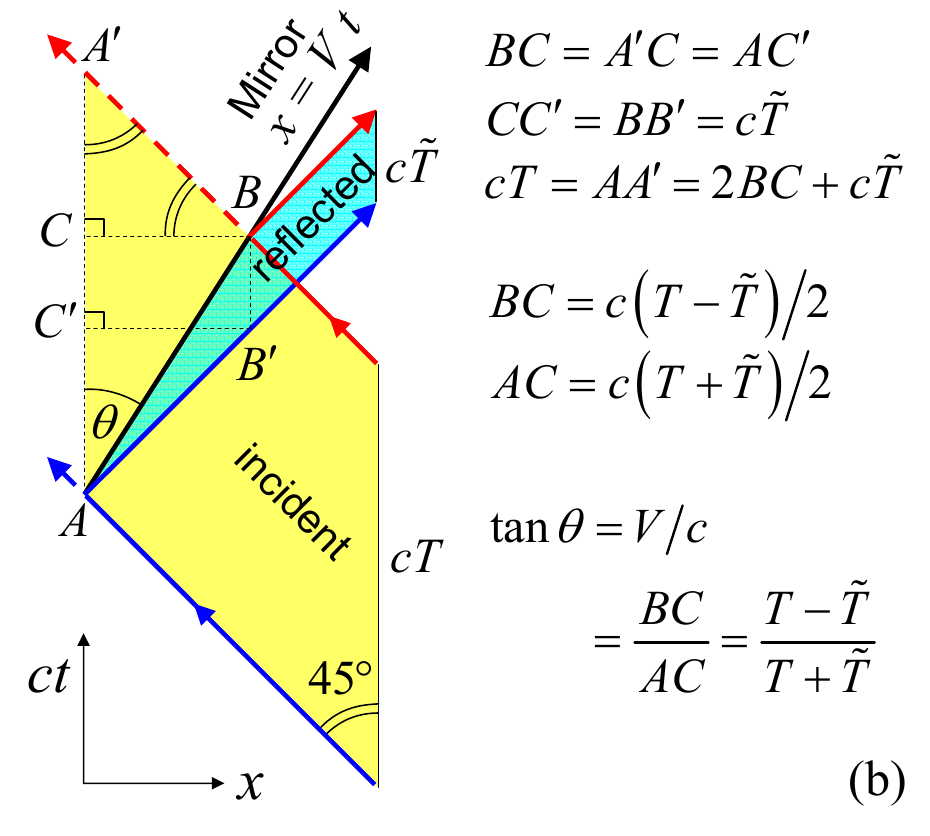}}
\caption{(a) The world lines of the incident, transmitted and
reflected electromagnetic pulses
and of the subluminal Mirror and superluminal Mirror-2.
(b) The geometric relashionship between durations
of the incident and reflected pulse sequences.
\label{fig:mirror}
}
\end{figure}

The factor by which the frequency
of the reflected electromagnetic radiation is multiplied
in the double Doppler effect
can be derived using simple geometric arguments, as shown
in Fig.~\ref{fig:mirror}.
The one-dimensional motion of an object
is represented by a world line in the plane ${x,ct}$,
where $x$ is the space coordinate, $t$ is time and $c$ is the speed of light in vacuum.
For an electromagnetic pulse
propagating in the direction of increasing $x$-values
the world line is stright and is inclined
at the angle of $+45^{\circ}$ (clockwise)
with respect to the $ct$-axis.
The world line of a counter-propagating pulse
is inclined at the angle of $-45^{\circ}$.
The world line of a mirror, moving with a constant velocity, $V$,
in the direction of the $x$-axis,
is also stright, but inclined at the angle of $\theta = \arctan(V/c)$.
In principle, the mirror velocity can be equal or greater
than the speed of light in vacuum,
e.~g., when the mirror is formed by an ionization front in plasma,
\cite{bib:Ostr}.
In Fig.~\ref{fig:mirror} (a) we show the world lines
for two semi-transparent Mirrors, one propagates with velocity $V<c$ and
another, superluminal, with velocity $V_2>c$.
We consider two consecutive electromagnetic pulses,
the "older" and the "younger", which interact with the mirrors.
The interaction is represented by the intersection
between the world lines of the pulses and Mirrors.
From the intersection points,
two reflected pulses are emitted.
We assume that the time of the formation of reflected pulses
is much less than the time period between these pulses.
Under this condition,
the time period, $\tilde T$, between the reflected pulses
is determined only by the time period, $T$, between the incident pulses
and the inclination angle of the world line representing the mirror.
In Fig.~\ref{fig:mirror} (b),
the points of intersection of the older and younger pulses
with the Mirror are denoted as $A$ and $B$, respectively.
From these points we draw two segments parallel to the $ct$-axis, $AA'$ and $BB'$,
where $A'$ lies on the world line of the younger pulse
and $B'$ lies on the wold line of the reflected older pulse.
The angle $A'AB$ (between the world line of the Mirror and the $ct$-axis)
is denoted as $\theta$. 
We draw two segments parallel to the $x$-axis, $BC$ and $B'C'$,
where $C$ and $C'$ lie on $AA'$.
By construction, $AA' = cT$ and $CC'=c\tilde T$.
Since the world lines of the electromagnetic pulses
form angles $\pm 45^{\circ}$ with respect to the $ct$-axis,
the triangles $A'BC$ and $AB'C'$ are isosceles and equal.
Therefore,
\begin{equation}
\tan\theta = \frac{T-\tilde T}{T+\tilde T}.
\end{equation}
On the other hand, it is equal to $V/c$.
Thus we obtain
\begin{equation}\label{eq:tt}
\tilde T = \frac{c-V}{c+V} T
\, .
\end{equation}
Assuming that the incident pulses in Fig.~\ref{fig:mirror}
represent two crests of an electromagnetic wave with frequency $\omega=2\pi/T$,
we obtain the reflected wave frequency, $\tilde \omega$:
\begin{equation}\label{eq:omega}
\tilde \omega  = \frac{c+V}{c-V} \omega
\, .
\end{equation}
In the limit $0<1-V/c \ll 1$, 
$\tilde \omega \approx 4\gamma^2 \omega$,
where
\begin{equation}
\gamma=\frac{1}{\sqrt{1-V^2/c^2}}
\end{equation}
is the Lorentz factor of the Mirror.
In the case of the superluminal Mirror-2,
$V>c$,
the time period between the reflected pulses
is given by Eq.~\ref{eq:tt} with a substitution $T\rightarrow T_2$.
However, as seen in Fig.~\ref{fig:mirror},
the wave reflected by a superluminal mirror is reversed in time.
We note that the presented derivation
of the frequency multiplication factor
does not use the Lorentz transformations;
in principle, it is valid for any sufficiently smooth world line
of the mirror provided that the act of reflection is sufficiently short
at each point of the mirror's world line.

%%%%%%%%%%%%%%%%%%%%%%%%%%%%%%%%%%%%%%%%%%%%%%%%%%%%%%%%%%
\section{The reflection coefficient of thin plasma slab}
\label{sec:coeff}

Here we recall the derivation of the reflection coefficient
of a moving thin plasma slab, following Refs. \cite{bib:FM,bib:Panchenko}.
The plasma slab with thickness, $l$, and density, $n_0$,
moves with velocity, $V$, along the $x$-axis, so that
the density as function of $x$ and $t$ reads
\begin{equation}
n(x-Vt) = \left\{
 \begin{array}{ll}
   n_0\, , & |x-Vt|\le l/2 \, , \\
   0 \, ,   & |x-Vt|>l/2 \, .
 \end{array}
\right.
\end{equation}
A counter-propagating electromagnetic wave,
represented by the $z$-component of the vector potential, $A_z (x,y,z,t)$,
is incident on the plasma slab.
In the approximation of a weak electromagnetic wave,
the interaction is described by the Maxwell equation:
\begin{equation}
\partial_{tt}A_{z}-c^{2}(\partial_{xx}A_{z}+\partial_{yy}A_{z})+
\frac{4\pi e^{2}n(x-Vt)}{m_{e}\gamma} A_{z}=0,  \label{wave-eq}
\end{equation}%
where $e$ and $m_e$ are the electron charge and mass,
$\gamma=(1-V^2/c^2)^{-1/2}$ is the Lorentz factor of the plasma slab.

We consider a thin slab approximation,
where the slab thickness tends to zero, $l\rightarrow 0$,
while the total number of particles in the slab is constant,
$n_0 l = {\rm const}$.
Such the density is described by the Dirac delta function,
\begin{equation}\label{dens-delta}
n({\mathrm{X}})  = n_{0} l \delta \left({\mathrm{X}}\right),
\end{equation}
where we introduce a new variable $\mathrm{X} = x-Vt$
(we note that $\delta(\mathrm{X})$ has the dimension of inverse length).
This approximation was used in Ref. \cite{bib:FM}
for calculating the reflection coefficient of the breaking wake wave,
basing on the argument that
in each period of the wake wave near breaking,
nearly half of the electrons are concentrated in the spike of
the electron density moving with the phase velocity of the wake,
whereas another half are distributed almost homogeneously
and moving in the opposite direction with the same velocity.

In the boosted reference frame, moving with the velocity of the plasma slab,
the electromagnetic wave is interacting with an infinitely thin wall at rest.
This allows us to use the analogy with the classical problem
of the scattering theory (e.g., see Refs. \cite{bib:Scott} and \cite{bib:FOIL}).
For the boosted frame,
we denote the spacial coordinate as
\begin{equation}
\xi =(x-Vt)\gamma,
\end{equation}
the time -- as $t'$,
and the frequency and wave number of the electromagnetic wave -- as
$\omega'$ and $\{ k'_x,k'_y,k'_z \} = \{ \kappa, k_y,k_z\}$.
Since the magnitude of the wave 4-vector
is zero in any inertial reference frame,
\begin{equation}
\kappa^2 = (\omega'/c)^{2}-k_{y}^{2}-k_{z}^{2}>0.
\end{equation}
We note that for $k_y=k_z=0$,
$\kappa = k_x (1+V/c)\gamma$.
Representing the wave vector-potential in the dimensionless form
\begin{equation}
a(\xi ) = \frac{eA_{z}(\xi )}{m_{e}c^{2}}\exp (-i(\omega't'- {k}_{y}{y}-{k}_{z}{z}))
\, ,
\end{equation}%
from Eq. (\ref{wave-eq}) and Eq. (\ref{dens-delta})
we obtain
\begin{equation}
\frac{d^{2}a(\xi )}{d\xi ^{2}}+
\left( \kappa^{2}-{\chi}\delta \left({\xi}\right)\right) a(\xi )=0
\, ,
\label{eqwedelta}
\end{equation}%
where
\begin{equation}
\chi= \frac{2\omega}{c} n_0 l r_e \lambda ,
\end{equation}
$\lambda=2\pi c/\omega$ is the wavelength
and $r_e = e^2 /m_e c^2$ is the classical electron radius.
Integrating Eq. (\ref{eqwedelta}) over $\xi$ in the interval $-\varepsilon <\xi<\varepsilon$,
in the limit $\varepsilon \rightarrow 0$
we obtain the boundary conditions for the right and left derivatives of $a(\xi )$: 
\begin{equation} \label{eqwedelta-der}
\left. \frac{da}{d\xi }\right \vert _{+0}-\left. \frac{da}{d\xi }%
\right \vert _{-0}=\chi a(0),
\end{equation}%
while $a(\xi )$ is continuous at $\xi=0$.
The solution to Eq. (\ref{eqwedelta}), describing the wave reflection,
is cast in the form
\begin{equation}
a(\xi) = a_0 \left\{
\begin{array}{ll}
e^{i \kappa \xi} + \rho(\kappa) e^{-i \kappa \xi}, & \xi\ge0, \\
\tau(\kappa) e^{i \kappa \xi}, & \xi<0,
\end{array}
\right.
\end{equation}
where $\rho$ and $\tau$ are related by
expressions following from the boundary condition:
\begin{eqnarray}
1+\rho (\kappa)=\tau (\kappa), \\
i\kappa(1-\rho (\kappa)-\tau (\kappa)) = \chi\tau (\kappa).
\end{eqnarray}%
This gives the amplitude of the reflected radiation
\begin{equation}
\rho (\kappa)=-\frac{\chi}{\chi+2i\kappa}=\tau (\kappa)-1.
\end{equation}
Correspondingly, the reflection coefficient
defined as $\mathsf{R} = |\rho(\kappa)|^2$,
is equal to
\begin{equation}
\mathsf{R} = \frac{\chi^2}{\chi^2 + 4\kappa^2}
=\frac{(n_0 l r_e \lambda)^2}{(n_0 l r_e \lambda)^2 +
(\omega'/\omega)^2 - c^2k_\perp^2/\omega^2}.
\end{equation}
We note that the reflection coefficient  is a Lorentz invariant,
since it is expressible via transverse components of the vector-potential,
which are invariant under the Lorentz transformations.
For the case of normal incidence,
$k_\perp^2=k_y^2+k_z^2=0$, we obtain
\begin{equation}
\mathsf{R} = \frac{(n_0 l r_e \lambda)^2}{(n_0 l r_e \lambda)^2 +
(c+V)/(c-V)}.
\end{equation}
If the plasma slab is sufficiently high,
$n_0 l r_e\lambda\ge 4\gamma^2$,
the reflection coefficient is close to 1.
In the opposite limit,
$n_0 l r_e\lambda\ll 4\gamma^2$,
we obtain
\begin{equation} \label{eq:Rmax}
\mathsf{R} \approx
\frac{(n_0 l r_e \lambda)^2}{4\gamma^2}
\, ,
\end{equation}
where we assume $\gamma\gg 1$.
We see that the reflection coefficient is proportional to the
square of the number of electrons in the slab,
i.~e. the reflection is coherent.
In the strongly nonlinear wake wave, where $\gamma = \gamma_{\rm ph} = \omega/\omega_{pe}$
and $l=2c\sqrt{2\gamma_{\rm ph} }/\omega_{pe}$,
Eq. (\ref{eq:Rmax}) gives
$\mathsf{R} = 2/\gamma_{\rm ph} ^{3}$ \cite{bib:FM}.
A systematic analysis of the wave-breaking regimes and
caustics formation as well as the derivation of the
reflection coefficients for a wide class
of the plasma density singularities
are presented in Ref. \cite{bib:Panchenko}.

%%%%%%%%%%%%%%%%%%%%%%%%%%%%%%%%%%%%%%%%%%%%%%%%%%%%%%%%%%
\section{The accelerating mirror}
\label{sec:AM}

\begin{figure}[ht]
\centerline{\includegraphics[scale=0.33]{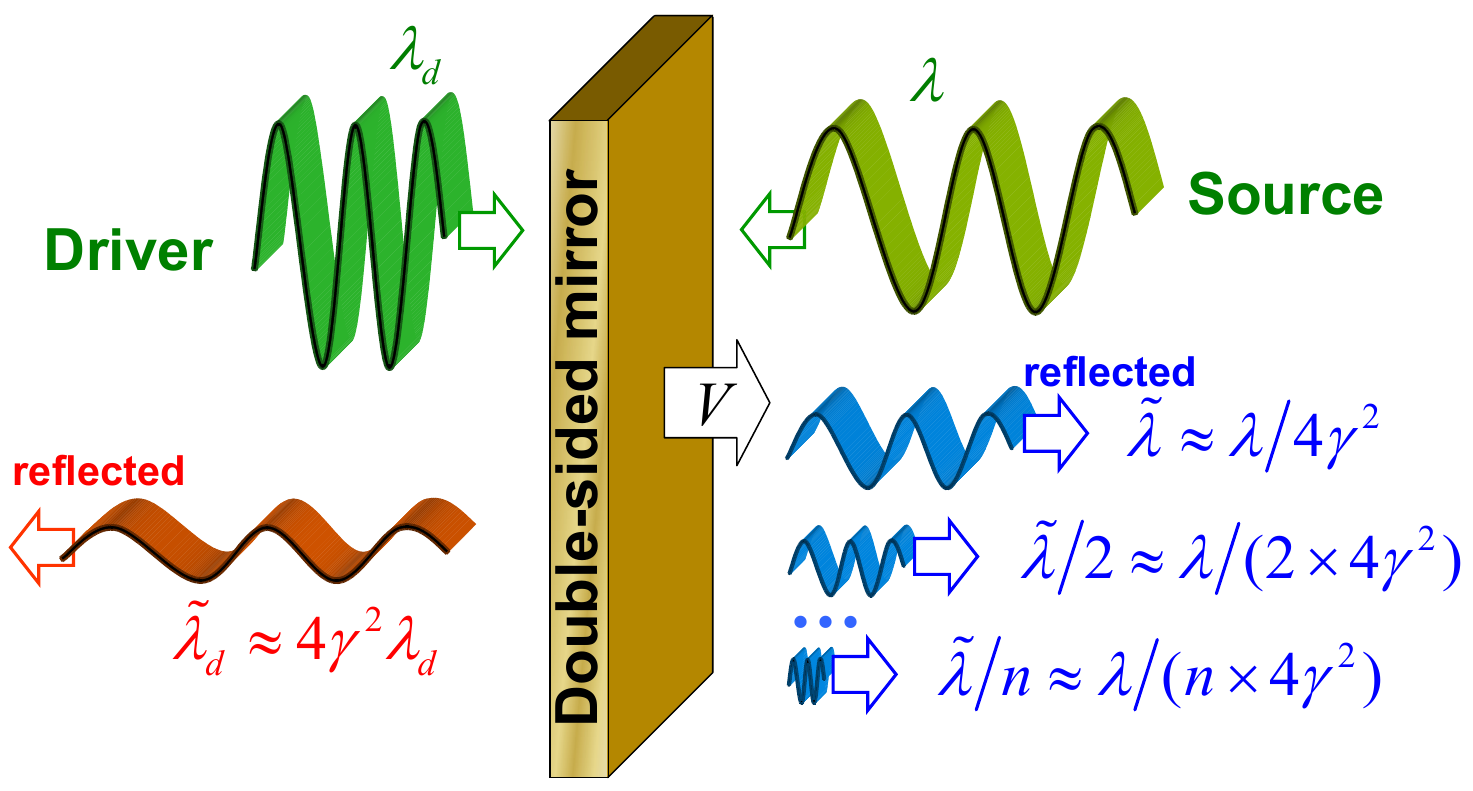}}
\caption{The scheme of the double-surface mirror.
The ultra-intense driver going from the left
accelerates the mirror in the radiation pressure dominant regime.
In its turn, the mirror reflects the 
intense source sent from the right.
The frequencies of generated high harmonics are also multiplied.
\label{fig:DSM}
}
\end{figure}

A plasma slab moving with relativistic velocity can be
created in the Radiation Pressure Dominant (RPD) regime of the ion acceleration
(also called the Laser Piston regime), \cite{bib:RPD}.
In this regime an ultra-intense electromagnetic wave (driver)
incident on a thin dense foil
efficiently accelerates the irradiated region of the foil
as a whole due to a strong radiation pressure and
high reflectivity of the foil, Fig.~\ref{fig:DSM}.
We consider the driver pulse 
with the electric field $E_L$ and length $L$,
carrying the energy ${\cal E}_L\propto E_L^2 L$.
The accelerated plasma slab, co-propagating with the driver,
forms a relativistic plasma mirror.
For simplicity we assume that it
perfectly reflects the driver pulse.
The length, $\widetilde{L}$, of the reflected pulse is longer
than that of the incident pulse by the factor $4\gamma^2$,
where $\gamma$ is the Lorentz factor of the plasma mirror.
The transverse electric field, $\widetilde{E}_L$, is smaller
than $E_L$ by the same factor.
Therefore,
after the reflection the driver energy becomes much lower:
$\widetilde{\cal E}_L\propto \widetilde{E}_L^2\widetilde{L} \approx
E_L^2 L/4\gamma^2$.
The mirror acquires the energy
$(1-1/4\gamma^2){\cal E}_L$ from the driver pulse.
The radiation momentum is transfered to ions
through the charge separation field
and the `longitudinal' kinetic energy of ions
is much greater than that of electrons.
According to Ref. \cite{bib:RPD},
the "plate" energy increases with time as $Nm_i \gamma \propto t^{1/3}$,
where $N$ is the number of ions with mass $m_i$.

If we send a counter-propagating relativistically strong electromagnetic wave (source)
onto the accelerated plasma slab,
it will be (partially) reflected
and the reflected wave (signal) frequency will be
multiplied due to the double Doppler effect, Fig.~\ref{fig:DSM}.
According to Eq. (\ref{eq:omega}),
the frequency of the reflected fundamental mode of the source
increases in time as
$\tilde\omega \approx 4\gamma^2 \omega \propto t^{2/3}$.
The plasma slab acts as a double-sided mirror, which
gain momentum reflecting the driver by one side
and transfer momentum to the electromagnetic wave
reflected by another side.
The source pulse should be sufficiently weaker than the driver,
nevertheless it can be relativistically strong.
In the interaction with a high-intense source pulse,
the accelerated plasma slab exhibits the properties
of the sliding and oscillating mirrors,
producing high harmonics.
As a result, in the spectrum of the reflected radiation
both the fundamental frequency of the incident radiation and all the high harmonics
are multiplied by the same factor, approximately
proportional to the square of the Lorentz factor of the mirror.

In order to investigate the feasibility of this effect
we performed two-dimensional particle-in-cell (PIC) simulations
using the Relativistic ElectroMagnetic Particle-mesh code REMP
based on the density decomposition scheme \cite{bib:REMP}.
The driver laser pulse with the wavelength $\lambda_d=\lambda$,
the intensity $I_d=1.2\times 10^{23}$W/cm$^2 \times (1\mu m/\lambda)^2$,
corresponding to the dimensionless amplitude $a=300$,
and the duration $\tau=20\pi c/\lambda$
is focused with the spot size of $D_d=10\lambda$
onto a hydrogen plasma slab with the thickness $l=0.25\lambda$
and the initial electron density $n_e=480n_{cr}
=5.4\times 10^{23}$cm$^{-3}\times (1\mu m/\lambda)^2$
placed at $x=10\lambda$.
The driver is $p$-polarized,
i.~e., its electric field is directed along the $y$-axis.
Its shape is Gaussian but whithout the leading part,
starting $5\lambda$ from the pulse center along the $x$-axis.
At the time when the driver pulse hits the plasma slab
from the left ($x<10\lambda$),
the source pulse arrives at another side of the slab from the right
($x>10.25\lambda$).
The source pulse is $s$-polarized (its electric field is
along the $z$-axis). It has the same wavelength as the driver pulse.
Its intensity is 
$I_s=1.2\times 10^{19}$W/cm$^2 \times (1\mu m/\lambda)^2$,
corresponding to the dimensionless amplitude $a=3$,
its duration is $\tau=120\pi c/\lambda$ and its waist size is $D_s=20\lambda$.
The source pulse has rectangular profile along the $x$- and $y$-axes;
such the profile is not necessary for the desired effect but helps to analyse the results.
We note that the $p$-polarization of the driver may be not optimal
for a smooth start of the slab acceleration in the
radiation pressure dominant regime, 
nevertheless it was chosen in order to easily distiguish between
the driver and the source pulses.
In addition, our choice demonstrates the
robustness of the double-surface mirror.

\begin{figure}[ht]
\centerline{\includegraphics[scale=0.4]{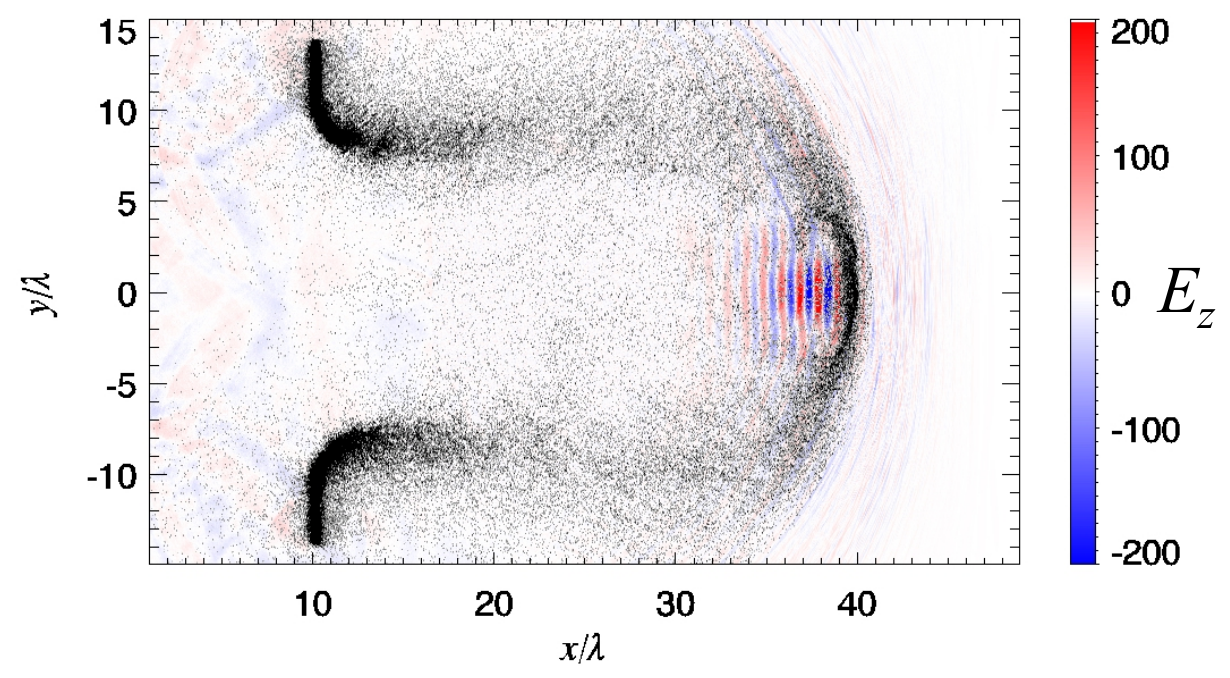}}
\caption{
The $y$-component of the electric field, representing the driver pulse,
and the ion density (black) at $t=37\times 2\pi/\omega$
after the driver has hit the plasma slab.
\label{fig:1}
}
\end{figure}

\begin{figure}[hb]
\centerline{\includegraphics[scale=0.3]{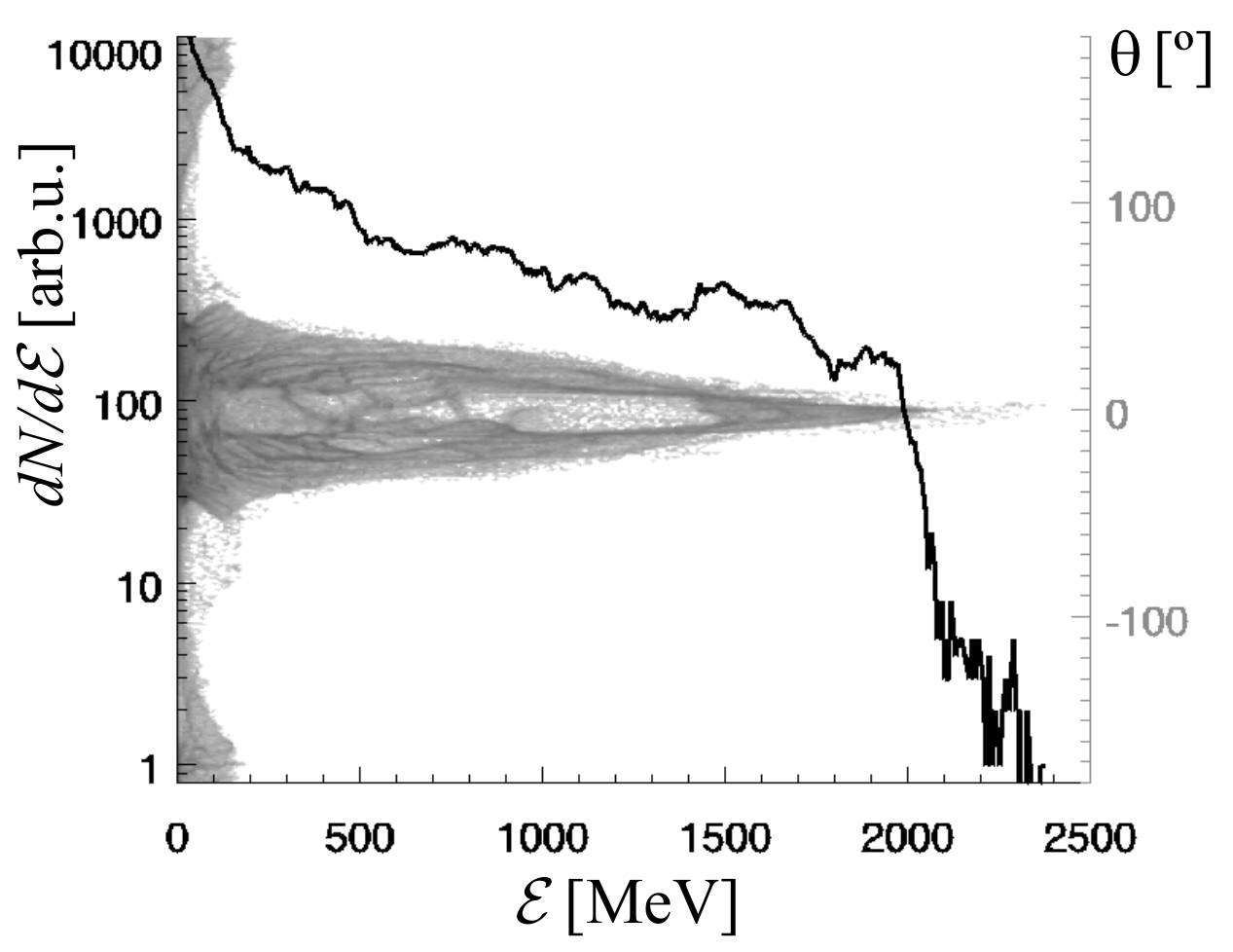}}
\caption{The energy (curve)
and the anglular (grayscale) distributions
of ions at $t=37\times 2\pi/\omega$.
\label{fig:2}}
\end{figure}

\begin{figure}[ht]
\centerline{\includegraphics[scale=0.4]{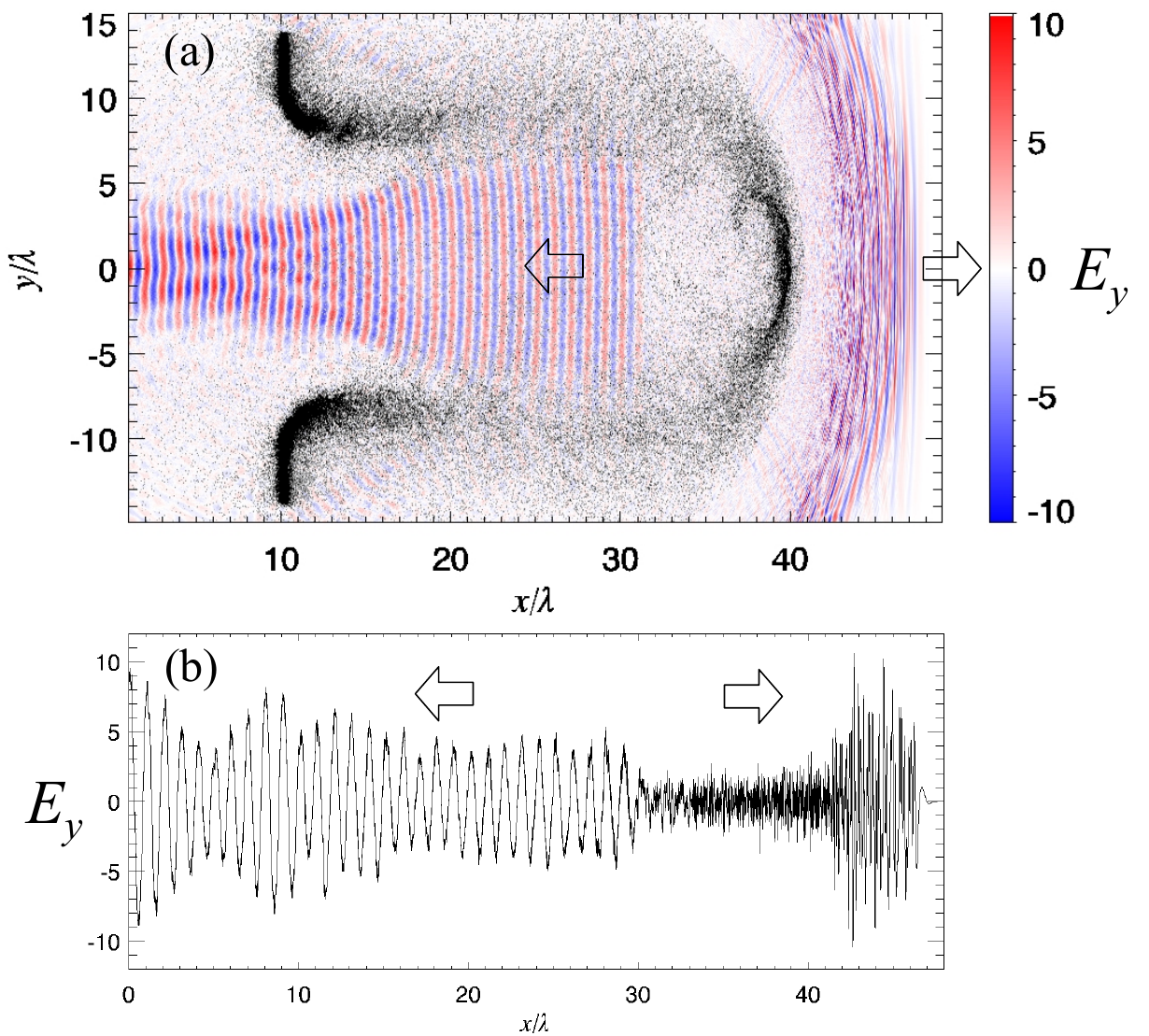}}
\caption{(a) The $z$-component of the electric field,
representing the source pulse,
and the ion density (black).
(b) The profile of the $z$-component of the electric field
along the $x$-axis at $y=0$.
Both the frames for $t=37\times 2\pi/\omega$.
In the frame (a), strongly jagged distribution is not seen
in the colorscale in the interval $40<x/\lambda<42$
due to the sampling (the wavelength is smaller than the image pixel).
\label{fig:3}
}
\end{figure}

\begin{figure}[hb]
\centerline{\includegraphics[scale=0.25]{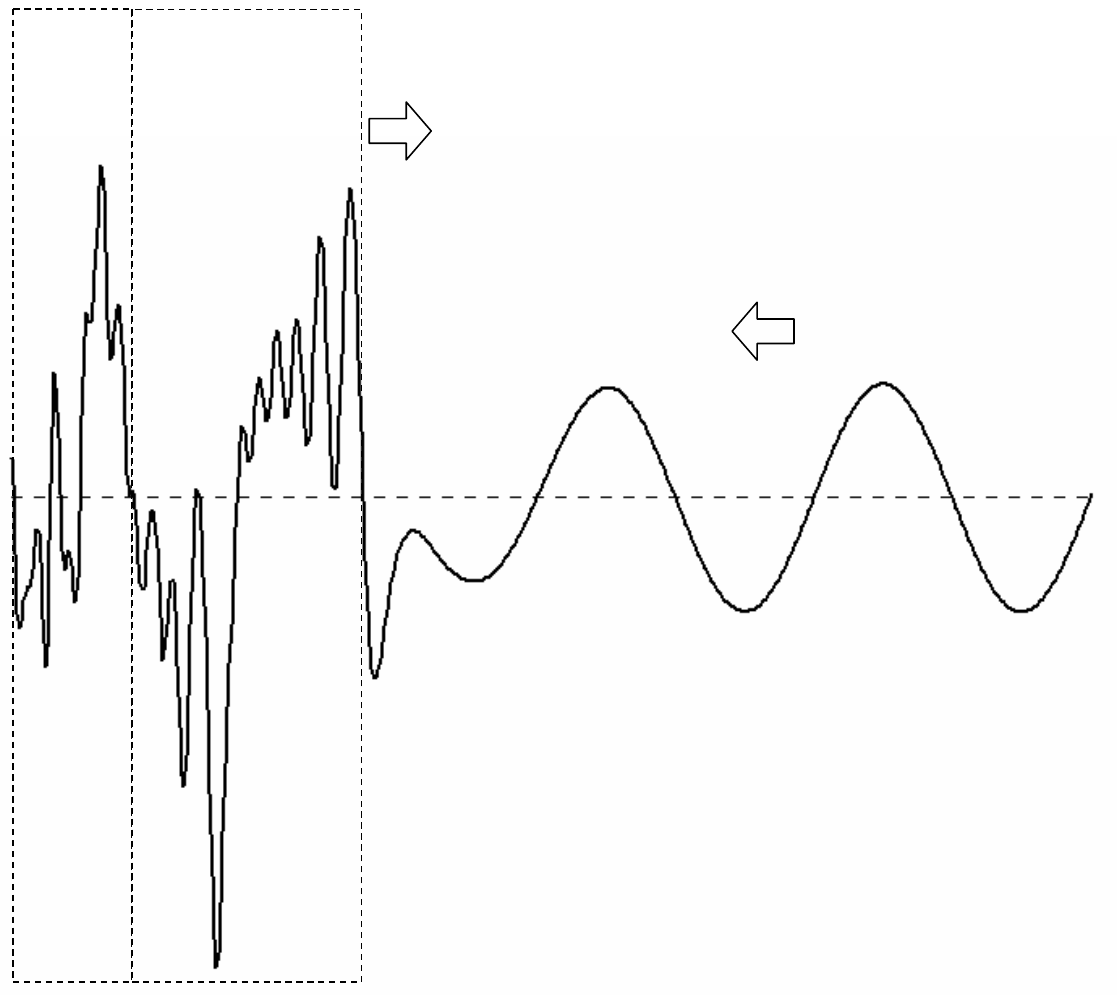}}
\caption{The $z$-component of the electric field along the $x$-axis
(at $y=0$), showing the incident source pulse overlapped with
the first two consecutive cycles of the reflected radiation.
Both the cycles exhibit presence of high harmonics.
The second cycle (left) is comressed in comparison with the first cycle
because the mirror velocity is increasing,
which results in the increase of the compression factor according to Eq. (\ref{eq:omega}).
\label{fig:4}
}
\end{figure}

The results of the simulations are shown in Figs. \ref{fig:1}-\ref{fig:4},
where the spatial coordinates and time units are in the laser wavelengths and
wave periods, respectively.
The accelerated portion of the plasma slab is seen in Fig. \ref{fig:1}.
The driver laser makes a cocoon where it is confined.
At the time of 37 laser periods from the beginning of the
driver-slab interaction, the ions are accelerated up to 2.4 GeV
while the majority of ions in the accelerated "plate" carry the energy 
about 1.5 GeV, Fig. \ref{fig:2}.
The source pulse is reflected from the accelerating plasma, as shown in Fig. \ref{fig:3}.
The frequency of the reflected radiation increases as the mirror moves faster,
thus the profile of the electric field along the axis becomes more and more jagged,
Fig. \ref{fig:3}(b).
In other words, the reflected pulse is chirped.
In addition, as seen in Fig. \ref{fig:3}, a portion of the source pulse
reflected from the curved edges of the expanding cocoon
acquires an inhomogeneous frequency upshift
determined by the angle of the reflecting region.
At the begining the magnitude of the reflected radiation is higher than
that of the incident source (3 times higher at maximum).
This is due to the frequency multiplication, specific to the double Doppler effect,
and due to compression of the plasma slab under the radiation pressure of the driver pulse.
Later the magnitude of the reflected radiation drops.
In an instant proper frame of the accelerating "plate",
where it is stationary for a given moment of time,
the frequency of the incident source electromagnetic wave
becomes higher with time, thus the plate becomes
more and more transparent in accordance with Eq. (\ref{eq:Rmax}).
Correspondingly, after some time the source starts to be transmitted through
the plasma more and more efficiently.
This is seen in Fig. \ref{fig:3} where the transmitted radiation
is focused because of the cocoon-like spatial distribution of the plasma.

The reflected radiation has a complex structure of the spectrum.
First, it contains not only the {frequen\-cy-multi\-plied}
fundamental mode of the source pulse, but also high harmonics due to the nonlinear
interaction of the source with the plasma slab. This is seen in Fig. \ref{fig:4},
showing first two consecutive cycles of the reflected radiation.
Both the cycles exhibit presence of high harmonics,
while the later cycle is compressed together with its harmonics
in comparison with the earlier cycle.
Second, the reflected radiation has a spectral shift due to the fact that
the electrons affording the reflection move along
the plasma slab under the action of the driver pulse.
Third, the spectrum is enriched by a continuois component
since the mirror moves with acceleration.
The theory of this spectral structure will be presented elsewhere.

%%%%%%%%%%%%%%%%%%%%%%%%%%%%%%%%%%%%%%%%%%%%%%%%%%%%%%%%%%
\section{Conclusion}
\label{sec:Conclusion}

In this paper we show that
a solid density plasma slab, accelerated in the radiation pressure dominant regime,
can efficiently reflect a counter-propagating relativistically strong laser pulse (source),
thus playing a role of accelerating mirror.
The reflected electromagnetic radiation 
consists of the reflected fundamental mode
and high harmonics, all multiplied by the factor $(1+V)/(1-V)\approx 4\gamma^2$,
where $V$ is the increasing velocity of the plasma slab and $\gamma$ is the corresponding
Lorentz factor.
In general, the reflected radiation is chirped due to the mirror acceleartion.
With a sufficiently short source pulse beeing sent with an appropriate delay to
the accelerating mirror, one can obtain a high-intese ultra-short pulse of X-rays.

For the mirror velocities greater than some threshold,
the distance between electrons in the plasma slab in the proper reference frame
becomes longer than the incident wavelength.
Thus the plasma slab will not be able to afford the reflection in a coherent manner,
where the reflected radiation power is proportional to the square of
the number of particles in the mirror.
In this case the reflected radiation becomes linearly proportional to the
number of particles.
Even with this scaling one can build an ultra-high power source of short gamma-ray pulses,
when the interaction of the source pulse with a solid-density plasma
is in the regime of the inverse Compton scattering.

Employing the concept of the accelerating high-density mirror,
one can develop a relatively compact and tunable ultra-bright high-power
X-ray or gamma-ray source, which will considerably expand the range of applications
of the present-day powerful sources and will create new applications and research fields.
Implementation in the "water window" will allow performing a single shot high contrast imaging of
biological objects. In atomic physics and spectroscopy, it will allow performing the multi-photon
ionization and producing high-Z hollow atoms. In material sciences, it will reveal novel properties
of matter exposed to the high power X-rays and gamma-rays.
In nuclear physics it will allow studying states of high-Z nucleus.
The sources of high-power coherent X-ray and ultra-bright gamma-ray radiation
also pave the way towards inducing and probing
the the nonlinear quantum electrodynamics processes.

%\begin{acknowledgement}
This work was partially supported by the Ministry of
Education, Science, Sports and Culture of Japan, Grant-in-Aid for Creative
Scientific Research No. 15002013.
The authors acknowledge the support by the
European Commission under contract ELI pp 212105
in the framework of the program FP7 Infrastructures-2007-1.
%\end{acknowledgement}

\end{document}